\documentclass[12pt,preprint]{aastex}

\slugcomment{Submitted to the Astrophysical Journal}

\shorttitle{The beaming pattern and spectrum 
of inverse  Compton radiation in blazars} 
\shortauthors{Georganopoulos, Kirk,\& Mastichiadis}

\newcommand{\egret}{{\it EGRET\/}}
\newcommand{\osse}{{\it OSSE\/}}
\newcommand{\comptel}{{\it COMPTEL\/}}

\begin{document}

\title{The beaming pattern and spectrum of radiation from
inverse Compton scattering in blazars}

\author{M. Georganopoulos, J.G. Kirk}
\affil{Max Planck Institut f\"{u}r Kernphysik, Postfach 10 39 80,  Heidelberg, D 69029, Germany}
\email{markos@mickey.mpi-hd.mpg.de,  John.Kirk@mpi-hd.mpg.de}
\and
\author{A. Mastichiadis}
\affil{Department of Physics, University of Athens, 
Panepistimiopolis,
GR 15783 Zografos, Greece}
\email{amastich@phys.uoa.gr} 
\begin{abstract}
By including Klein--Nishina effects,
we generalize previous calculations of the 
beaming pattern of photons produced by inverse Compton scattering.
For an isotropic distribution of soft photons upscattered by  
nonthermal electrons with a power-law density distribution
$n(\gamma)\propto \gamma^{-p}$, embedded 
in a plasma moving with relativistic bulk speed we show that 
the observed radiation intensity is proportional to 
${\cal D} ^{3+p}$, 
where ${\cal D}$ is the Doppler boosting factor. 
This agrees with previous 
computations performed in the Thomson limit, where
the observed spectral index is $\alpha=(p-1)/2$
and the beaming pattern ${\cal D}^{4+2\alpha}$.
Independent of ${\cal D}$, Klein--Nishina effects limit the 
location of the peak energy $\epsilon_{peak}m_ec^2$ of the observed spectral 
energy distribution such that $\epsilon_{peak} \lesssim  1/\epsilon_0$, 
where $\epsilon_0$ is the energy of the seed photons
in units of $m_e c^2$.
Assuming that the seed photons originate in the broad line region,
we demonstrate that the GeV emission of blazars
is  significantly modified  by Klein--Nishina 
effects, the spectrum being softer than that calculated in the Thomson limit. 
We further show that the change in 
spectral index of the inverse Compton emission
across $\epsilon_{peak}$ can exceed the value of $0.5$
predicted by computations performed in the Thomson limit. 
The model spectra agree with \osse\ and \comptel\
limits on this break 
without invoking the effects of differential absorption 
at the edge of a gamma-ray photosphere.
\end{abstract}

\section{Introduction}

Inverse Compton scattering, a process in which a nonthermal distribution of electrons boosts the energy of a \lq seed\rq\ or \lq target\rq\ photon by a large factor, is commonly thought to be responsible for the production of gamma-ray photons in blazars \citep{ulrichmaraschiurry97}.
For very high energy electrons, the
scattered photon can carry off a substantial fraction of the energy of the 
incoming electron.
There have been
several suggestions concerning the source of the target photons: 
 optical/UV 
photons from an accretion disk \citep{dermer92, boettchermauseschlickeiser97},  optical photons from the 
broad line region and infrared dust photons \citep{sikora94, blazejowski00}. 
The radiation mechanism for such targets is usually called external
Comptonization (\lq EC\rq).
Synchrotron photons produced in the jet can also act as seed photons for 
inverse Compton scattering, a process referred to  as synchrotron self-Compton
(\lq SSC\rq) scattering \citep{maraschi92, bloom96, inouetakahara96, mastichiadis97}.
The same synchrotron photons can serve as seed photons after being reflected 
back into the jet from  the broad line clouds \citep{ghisellini96}.
In those blazars that have strong emission lines, the
\egret-detected GeV emission is probably 
dominated by inverse Compton
scattering of the broad line photons \citep{sikora94, sikoraetal97}. 
In this case, the plasma
responsible for the GeV emission  moves relativistically 
with a bulk Lorentz factor $\Gamma$ through an approximately isotropic 
photon field.

For such a \lq blob\rq\ 
of plasma moving with 
velocity $\beta c=c(1-\Gamma^{-2})^{-1/2}$, at an 
angle $\theta$ to the observer's line of sight,
the beaming pattern of the observed power per unit solid angle
and per unit frequency is ${\cal D}^{3+\alpha}$ for any process in which the 
emission is isotropic in the 
frame
comoving with the blob (the \lq blob frame\rq).
Here, 
${\cal D}=1/[\Gamma(1-\beta\mu)]$ is the familiar Doppler factor 
with $\mu=\cos\theta$, and $\alpha$ is the spectral index of the 
radiation. An isotropic electron distribution and a tangled magnetic 
field, produce, in the 
blob frame, isotropic 
synchrotron and SSC emission, which is,  
therefore, characterized by this beaming factor
\citep{begelman84}. The situation is different for inverse Compton 
scattering of photons on targets which are approximately isotropic not in the
blob frame, but in the rest frame of the broad line region 
(\lq lab.\ frame\rq).
In the Thomson limit, \cite{dermer95} has shown that the beaming pattern
in this case is ${\cal D}^{4+2\alpha}$. In terms of the power-law index of the
electron distribution, this is equivalent to ${\cal D}^{3+p}$. 
In \S~\ref{math} we show that this 
expression arises immediately from the transformation properties of the
electron distribution and is valid for arbitrary electron energy. Thus, in the
case of EC emission, the beaming is most simply expressed in terms of 
the underlying electron
distribution, whereas for SSC and synchrotron emission 
it can be obtained straightforwardly from the observed
photon distribution.
In addition to the beaming pattern for a power-law distribution,
\citet{dermer95} also demonstrated that a characteristic energy, such as
a cut-off or break in the spectrum, is observed
at a photon energy that
scales as 
$\epsilon_{max}\propto {\cal D}^2$. This contrasts with the 
situation for synchrotron and SSC radiation, 
where the scaling is linear with ${\cal D}$. Once again, the
difference can be understood from the transformation  properties of electrons
and photons, respectively. We show that the scaling changes when
Klein-Nishina (\lq KN\rq ) effects intervene: for hard electron distributions
 ($p<3$) with an upper cut-off, 
the peak in the spectral energy distribution is limited to a value 
$\epsilon_{peak} \lesssim 1/\epsilon_0$ that is independent of ${\cal D}$, 
where $\epsilon_0$ is the energy of the seed photons
in units of $m_e c^2$. (Energy units
of $m_e c^2$ are used throughout this work.)

Assuming that the \egret-detected GeV emission from blazars is due to 
inverse Compton scattering of broad line region photons we show in \S~3
that the
spectra are significantly affected by KN effects, which 
result in a softening of the \egret\ 
spectrum. This effect has been noted in connection
with blazar spectra by \citet{boettchermauseschlickeiser97},
who computed model spectra using anisotropically distributed accretion 
disk photons as targets, rather than photons
from the broad line region.  
 We map the observed \egret\ spectral indices 
to the index of the electron distribution, and find
that these are grouped around the value $p=3.2$, in agreement
with recent theoretical work on particle acceleration in relativistic shocks
\citep{bednarzostrowski98,kirk00,achterbergetal01}. 
Synchrotron cooling or, alternatively, cooling by IC scattering in the Thomson
limit produces a break of $\Delta p=1 $ in the electron spectrum. 
The spectrum of IC radiation in the Thomson limit therefore shows a break of
$\Delta \alpha=0.5$. However, some $\gamma$-ray sources that peak at MeV 
energies show a more pronounced break \citep{mcnaron95}, which has been
interpreted as evidence for gamma-ray absorption by pair production
\citep{blandfordlevinson95,marcowithhenripelletier95}. 
We show that spectral
softening can explain these observations, provided the electron distribution
is determined by synchrotron cooling.
A summary of our results and their implications is presented in 
\S~\ref{discussion}.

\section{The beaming pattern of inverse Compton radiation \label{math}}

Consider a blob of plasma moving  relativistically with a bulk Lorentz factor
$\Gamma$ and velocity $\beta c$, at an angle $\theta$ to the observer's line of sight.
In the frame of the blob the electrons are characterized
by an isotropic power-law density distribution $n'(\gamma')$,

\begin{equation}
n'(\gamma')={ k \over 4\pi} \gamma'^{-p}\,P(\gamma_1,\gamma_2,\gamma'),
\end{equation}  
where $\gamma'$ is the Lorentz factor of the electron, $k$ is a constant, and
$P(\gamma_1,\gamma_2,\gamma)=1$ for $\gamma_{1}\leq \gamma\leq \gamma_2$, and
zero otherwise.
Under the assumption that $\gamma' \gg \Gamma$, one can treat the electrons
as a photon gas and make use of the Lorentz invariant quantity $n/\gamma^2$.
 The Lorentz factor $\gamma$ of an electron in the
lab. frame is then  $\gamma={\cal D} \gamma'$ 
and the electron density $n(\gamma)$ in the lab. frame is 
\begin{equation}
n(\gamma,\mu)=n'(\gamma')\left( {\gamma \over \gamma'}\right)^2={k \over 4\pi}{\cal D}^{2+p} \gamma^{-p}\,P(\gamma_1{\cal D},\gamma_2{\cal D},\gamma).  
\end{equation} 
Given that the effective volume $V_{\rm eff}$ of the blob is $ V_{\rm eff}=V{\cal D}$ (see Appendix A), where 
$V$ is the volume of the blob in the blob frame, the energy 
distribution of the effective  number of electrons $N_{\rm eff}(\gamma,\mu)$ is:
\begin{equation}
N_{\rm eff}(\gamma,\mu)= n(\gamma,\mu)V_{\rm eff}=\displaystyle{kV \over 4\pi}{\cal D}^{3+p} \gamma^{-p}\,P(\gamma_1{\cal D},\gamma_2{\cal D},\gamma). 
\label{eq:ele} 
\end{equation} 

Consider now that this electron distribution will Compton--scatter  
seed photons of  an arbitrary angular distribution. 
Since $\gamma\gg 1$, we can make the approximation usually made in synchrotron 
theory that the outgoing  photons move in the same direction as the scattering
electron. Therefore, when observing under
a certain angle only electrons moving in this direction  contribute to 
the Compton luminosity. Since the effective number of these electrons is
proportional to    
${\cal D}^{3+p}$, the  Compton specific luminosity  (observed luminosity  per energy interval  per solid angle) is  also proportional to 
${\cal D}^{3+p}$. Different seed 
photon angular 
distributions will  introduce  an angle--dependent  multiplication 
term in the calculation of the external Compton luminosity. 
An extreme case of a photon angular distribution 
is a  monodirectional photon beam propagating along the  direction of 
motion of the blob, for which 
\citet{dermer92} calculated the 
beaming factor of Compton scattering to be ${\cal D}^{3+p}  (1-\mu)^{(p+1)/2}$. Note here that for $\theta=0^o$ the inverse Compton luminosity is zero.
It can easily be shown that for a monodirectional photon beam propagating
in the direction opposite to the  direction of 
motion of the blob, as may be the case for the mirror model of \citet{ghisellini96}, the beaming factor is ${\cal D}^{3+p}  (1+\mu)^{(p+1)/2}$.

We now consider that the plasma blob is propagating through 
an environment permeated  by an isotropic monoenergetic photon field of
 energy density $U$.
The lab. frame rate  of scatterings per final photon 
energy interval  for an electron of
Lorentz factor $\gamma$ is:
\begin{equation}
{dN_p \over dt d\epsilon}={3 \sigma_T c \over 4 \epsilon_0 \gamma^2} f(x), 
\label{eq:sc_rate}
\end{equation} 
where $\sigma_T$ is the Thomson cross section. \citet{jones68} introduced  the \lq head-on\rq\ 
 approximation in which the seed photons are treated as coming from the 
direction opposite to the electron velocity. Using this, which is valid for
for $\gamma \gg 1$, and  the full KN
cross-section for inelastic Compton scattering, he showed that
\begin{eqnarray}
& \displaystyle f(x)=\left[2x \ln x+x+1-2x^2+{(4\epsilon_0\gamma x)^2\over 2( 1+4 \epsilon_0 \gamma x)}\right] \,P(1/4\gamma^2,1,x)     , \\ \nonumber & \displaystyle x= { \epsilon \over 4 \epsilon_0 \gamma^2 (1- \displaystyle{ \epsilon \over \gamma })}.
\end{eqnarray} 
The maximum observed energy is
\begin{equation}
\epsilon_{max,KN}={4\epsilon_0 \gamma_2^2{\cal D}^2\over(1+4\epsilon_0\gamma_2{\cal D})}.
\label{eq:gmaxkn}
\end{equation}
In the  case  of Thomson   scattering ($\gamma\epsilon_0 \ll 1$), 
 \citet{rl79}, assuming isotropic scattering
in the electron  frame,  showed that
\begin{equation}
f(x)={2 \over 3} (1-x)\,P(1/4\gamma^2,1,x) ,\;\;\;\; x= {\epsilon \over 4\gamma^2\epsilon_0},
\label{eq:thomson}
\end{equation}
and that the maximum observed final energy is
$\epsilon_{max,T}=4\epsilon_0\gamma_2^2{\cal D}^2$. 

We now make the approximation that the outgoing photons are directed 
along the direction of the scattering electrons, which is justified provided
the electron angular distribution varies slowly over angular scales 
$\lesssim 1/\gamma$.
To obtain the specific luminosity one 
integrates the scattering rate (\ref{eq:sc_rate}) 
over the electron energy distribution (\ref{eq:ele}), and multiplies
the result by the observed photon energy $\epsilon m_e c^2$ and by the
photon number density $n_p=U/\epsilon_0 m_e c^2$

\begin{equation}
{dL \over  d\epsilon d\Omega}={\cal D}^{3+p} \;{ 3 kV \sigma_T c U \over 16 \pi \epsilon_0}\; {\epsilon \over \epsilon_0}\int^{\infty}_{1}\gamma^{-(2+p)} f(x)P(\gamma_1{\cal D},\gamma_2{\cal D},\gamma) d\gamma .
\label{eq:sc_int}
\end{equation} 
In the Thomson case, for energies $\epsilon_{min,T}\leq\epsilon\leq\epsilon_{max,T}$, where 
$\epsilon_{min,T}=4\epsilon_0\gamma_1^2{\cal D}^2$, the lower limit 
of the integration in equation (\ref{eq:sc_int}) is $\gamma_{min}=(\epsilon/4\epsilon_0)^{1/2}$, and the upper limit is $\gamma_{max}=\gamma_2{\cal D}$.  Performing
the elementary integral using  equation (\ref{eq:thomson}) we obtain:

\begin{eqnarray}
& \displaystyle {dL \over  d\epsilon d\Omega}={\cal D}^{3+p} \;{ kV \sigma_T c U \over 8 \pi \epsilon_0}\; {\epsilon \over \epsilon_0} \times \\  \nonumber                        &  \displaystyle   \left[(\gamma_2{\cal D})^{-(1+p)} \left({\epsilon\over 4\epsilon_0(3+p)(\gamma_2{\cal D})^2 } -{1 \over 1+p}  \right) \right. \\ \nonumber & \displaystyle \left. + \left( {\epsilon \over 4\epsilon_0}\right)^{-(1+p)/2}{2 \over (1+p)(3+p)} \right].  
\label{thomsonolo}
\end{eqnarray} 
For $p > -1$ and $\epsilon \ll \epsilon_{max,T}$ we have $\gamma_{min} \ll \gamma_{max}$.
Since the integrand is then steeper than $\gamma^{-1}$,
the above result  simplifies to
\begin{equation}
{dL \over  d\epsilon d\Omega}\approx{\cal D}^{3+p}                                      { kV \sigma_T c U 2^{p-1} \over  \pi \epsilon_0 (1+p)(3+p)}                                                                 \left( {\epsilon \over \epsilon_0}\right)^{-(p-1)/2}.  
\label{thomson}
\end{equation}
The beaming of the observed radiation is the direct outcome of the electron
beaming, and it is characterized by the electron index $p$.
In the Thomson limit, 
the resulting spectrum is a simple power law with a spectral index 
$\alpha=(p-1)/2$ and one can substitute for $p$ in 
equation (\ref{thomson})
 to recover  the ${\cal D}^{4+2\alpha}$ beaming 
pattern (equation~7 of D95).  
The  spurious term $(1+\mu)^{(\alpha+1)}$ in the result of D95, which, however, varies only slowly with $\mu$ for viewing angles of interest, was introduced
by the approximation that the seed photons in the  frame of the blob are 
coming from a direction opposite to the  direction of the velocity of the blob.
In both  D95 and here the maximum observed energy 
$\epsilon_{max,T}$, as well as any other characteristic energy, scale as 
$\propto {\cal D}^2$, whereas in synchrotron and SSC they scale as ${\cal D}$.
If, instead of observing at a fixed energy, 
we are interested in the specific luminosity measured at a break or cut-off in
the spectrum, then the ${\cal D}^2$ scaling of the break energy introduces an
additional ${\cal D}^{-2\alpha}$ factor, so that the specific luminosity at the
break scales as ${\cal D}^4$.
The  luminosity per logarithmic energy interval of the spectral feature,
given by $\epsilon \, dL/d\epsilon d\Omega$, then  scales  as ${\cal D}^6$, 
since $\epsilon\propto{\cal D}^2$.

In the KN case, for energies $\epsilon_{min,KN}\leq\epsilon\leq\epsilon_{max,KN}$, where 
$\epsilon_{min,KN}=4\epsilon_0\gamma_1^2{\cal D}^2/(1+4\epsilon_0\gamma_1{\cal D})$, the lower limit 
of integration in equation (\ref{eq:sc_int}) is 
found by setting $x=1$
\begin{equation}
\gamma_{min}={\epsilon \epsilon_0 + \sqrt{\epsilon^2 \epsilon_0^2 + \epsilon \epsilon_0} \over 2\epsilon_0}.
\end{equation}
In this case the integrand is also steeper that $\gamma^{-1}$, 
and for $\gamma_{min} \ll \gamma_2{\cal D} \Rightarrow \epsilon \ll \epsilon_{max,KN}$, the
integration is dominated by the lower limit $\gamma_{min}$ which is independent
of ${\cal D}$. Therefore, the beaming pattern  ${\cal D}^{3+p}$ is also 
valid in the general case of KN scattering. 
The maximum energy is given by equation (\ref{eq:gmaxkn}), which 
is reduced to $\epsilon_{max,KN}=\gamma_2{\cal D}$ when the high energy tail
of the electron energy distribution is well into the KN regime, 
$\gamma_2 {\cal D} \epsilon_0 \gg 1$, a behavior similar to that of 
synchrotron and SSC emission 
maximum observed energies.
For electron indices $p < 3 $, $\epsilon_{peak,KN}$ cannot exceed significantly
the energy at which KN effects become important 
 and the scattering cannot be considered
elastic. For a given seed photon energy $\epsilon_0$ this sets in for 
electrons with energies $\gamma\approx 1/\epsilon_0{\cal D}$. Setting this limiting value
of $\gamma$ in equation (\ref{eq:gmaxkn})
 we obtain $\epsilon_{peak,KN}\lesssim 1/\epsilon_0$, independent of ${\cal D}$
and $\gamma_2$, provided the system is well into the KN regime, 
$\gamma_2 {\cal D} \epsilon_0 \gg 1$. 

\begin{figure}
\epsscale{0.75}
\plotone{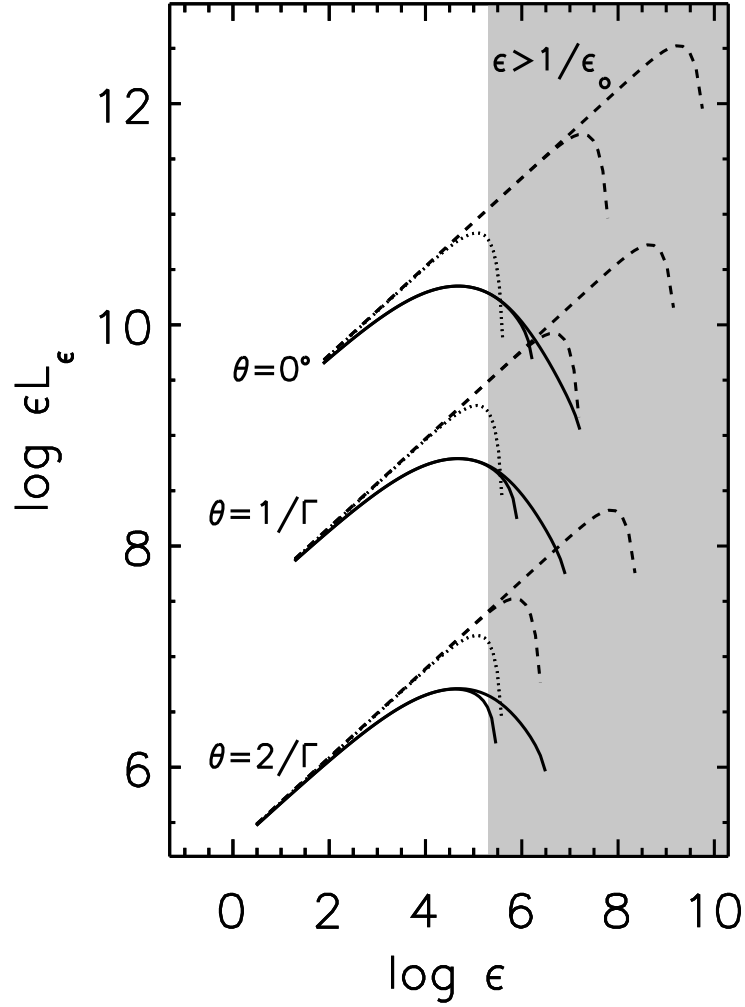}
\caption{. The observed  energy distribution due to inverse Compton 
scattering as a function of different  observing angles for  
the KN (solid lines),
the  Thomson (broken lines), and the  $ \left[ \; \sigma=\sigma_T  \right.$ for $ \gamma {\cal D} \epsilon_0< 3/4$, $0$ otherwise$\left. \right]$ (dotted lines) cross sections  for  a blob of plasma  that moves with a 
Lorentz factor $\Gamma=10$ through an isotropic monoenergetic photon field.
 The seed photon energy is $\epsilon_0=5 \times 10^{-6} $ in units of 
$m_ec^2$, 
which corresponds to optical photons. 
The electrons in the blob frame are characterized by an isotropic power law distribution
$n(\gamma)\propto \gamma^{-p},\;p=2.2,\;\gamma_1 \leq \gamma\leq \gamma_2$.
  For  each case we plot 
in normalized units the result for 
both $\gamma_2=10^5$ and $\gamma_2=10^6$, with the line corresponding to the
higher $\gamma_2$ reaching higher photon energies.}
\label{fig1} 
\end{figure}

We demonstrate these points in   figure \ref{fig1}, where  
we  plot the inverse Compton  spectral energy distribution for 
three different observing angles for both the Thomson
and KN cases and for two different values of $\gamma_2$. 
We also plot the 
spectral energy distribution calculated using the approximation  $ \left[ \; \sigma=\sigma_T  \right.$ for $ \gamma {\cal D} \epsilon_0< 3/4$, $0$ otherwise$\left. \right]$ (e.g. Chiaberge and Ghisellini 1999, Blazejowski et al. 2000).  
In the Thomson case
we  use the 
analytical expression (\ref{thomsonolo}), while in  the KN case we perform the
integration in equation (\ref{eq:sc_int}) numerically. 
The Thomson and the Klein Nishina  distributions deviate from each 
other with the 
KN spectrum being softer.
Note that  the deviation is already
significant at $\epsilon\approx 10^4$, which corresponds approximately to
electrons with Lorentz factor 
$\gamma {\cal D}\approx(\epsilon/\epsilon_0)^{1/2}\approx 4 \times 10^4$ in the lab. frame.
Therefore, already at $\gamma {\cal D} \epsilon_0 \approx 0.2 $, the Thomson
description is inadequate, and the  KN formalism must be used. 
Both the maximum and  peak energy of the Thomson spectral energy distribution
scale as $(\gamma_2{\cal D})^2$.
Contrary to this behavior, in the KN case the maximum energy scales as 
$\gamma_2{\cal D}$, whereas 
 the peak energy is insensitive to
variations of both ${\cal D}$ and $\gamma_2$ and it is located at an energy 
$\epsilon_{peak,KN}\lesssim 1/\epsilon_0$. The exact value of $\epsilon_{peak,KN} $ is a function of the electron index $p$, with steeper electron power laws
being characterized by lower $\epsilon_{peak,KN} $  values.
An increase in the upper 
cut-off $\gamma_2$ of the electron distribution by a factor of 10
affects only the steep high energy tail of the observed KN spectral 
energy distribution, leaving the peak energy and the peak
luminosity  unchanged. In general, as long as the scattering is KN limited (
$\gamma_2{\cal D}\epsilon_0 \gg 1$) , the peak energy will be insensitive to 
variations of both $\gamma_2$ and ${\cal D}$, in contrast to the Thomson
calculation and the synchrotron and SSC cases. 
The result  based on the step function cross section 
 $ \left[ \; \sigma=\sigma_T  \right.$ for $ \gamma {\cal D} 
\epsilon_0< 3/4$, $0$ otherwise$\left. \right]$ is practically identical to 
 the Thomson one, up to the cutoff energy $1/\epsilon_0$. 
In the case of  external Compton scattering of optical-UV  photons in 
blazars, where   $1/\epsilon_0\approx 100$ GeV,  the spectrum calculated 
under this approximation in the {\it EGRET} 
regime is for practical purposes the same as that calculated in 
 the Thomson regime.

As shown in figure~\ref{fig1}, the KN spectral energy distribution 
resulting from a power law electron energy distribution is not a power law, 
and one cannot assign
a unique spectral index to it. The beaming pattern
at a given energy  is expressed through the electron index $p$ and 
is not a simple function of the local spectral
index --- different parts of the spectrum
have the same beaming pattern ${\cal D}^{3+p}$
independent of the local spectral index. 
This is in contrast to  
synchrotron and SSC emission from a power law electron distribution, 
where the beaming pattern depends only on the
local spectral index, the steeper parts of the spectrum having a more
pronounced beaming behavior.

\section{Applications to blazars \label{blazars}}

The spectra of the \egret-detected blazars
\citep{hartman99} are described by simple power laws over the energy range
$30$ MeV -- $10$ GeV with no indication of a cut-off at high energy.
The photon indices of those \egret-detected blazars 
that display strong emission lines
cluster around $\approx 2.2 $ \citep{mukherjee97}, indicating that
the peak energy of the $\gamma$--ray spectral energy distribution in general
lies at energies below the \egret\ range. 
X-ray \citep{kubo98} \osse, and  \comptel\ observations 
\citep{mcnaron95} confine this peak to 
between about $1$ and $100$~MeV.

If  the observed radiation is inverse Compton emission from
optical/UV broad line seed photons ($\epsilon_0\approx  10^{-5}$),
 a  peak at $\approx 10 $ MeV arises 
from electrons with $\gamma{\cal D}\approx 1.5 \times 10^3$.
Since  
$\epsilon_0\gamma{\cal D} \approx 1.5 \times 10^{-2}$, 
the scattering can be adequately approximated by elastic Thomson scattering. 
Therefore, the peak at $\approx 10$ MeV  is not connected to KN effects and
must  result from a break
in  the electron energy distribution.
On the other hand, the \egret-observed flux is KN affected, and cannot be
described by elastic Thomson scattering. The 2 GeV flux ($\epsilon\approx 
 4 \times 10^3$) results from electrons with $\gamma{\cal D}\approx  2\times 10^4$, corresponding to $\epsilon_0\gamma{\cal D} \approx  0.2$, a regime in 
which the KN steepening of the spectrum relative to the Thomson case 
 is significant. 

In  models in which particle acceleration competes with radiative losses and
 particle escape from the system, the electron energy distribution is 
characterized by $\gamma_0$, the Lorentz factor 
at which electrons are injected,
$\gamma_b$, the electron Lorentz factor  at which the radiative 
cooling time equals to the escape time, and $\gamma_{max}$, 
the  electron Lorentz factor at which the acceleration time equals the 
radiative
cooling time.  Between $\gamma_0$ and $\gamma_b$ the electron distribution
is a power law with index $p$, while above $ \gamma_b $
the index steepens to  $p+1$, in the case where synchrotron cooling
dominates.
When  the electron index $p\leq2$, the peak of the inverse
 Compton  spectral energy distribution  is due to electrons 
with $\gamma\approx \gamma_{max}$ in the Thomson case, 
and the spectrum after the 
peak is  expected to decrease abruptly. On the other hand, when 
the electron index $p\geq2$, the peak of the inverse
 Compton  spectral energy distribution  is due to electrons 
with $\gamma\approx \gamma_{b}$ in the Thomson case, 
and the spectrum after the 
peak is  expected to follow a power-law behavior, up to a cut-off energy
associated with $\gamma_{max}$. 
In the case of the \egret-detected blazars 
the fact that after the Thomson dominated peak at $\approx 10 $ MeV there is a power law 
extension of the emission  at least up to $\approx 10$ GeV indicates that the 
 peak of the spectral energy distribution
is  associated with  electrons at $\gamma=\gamma_b$
where the radiative cooling time equals the escape time from the system,
and that the electron index $p \geq 2$.

\begin{figure}
\epsscale{0.75}
\plotone{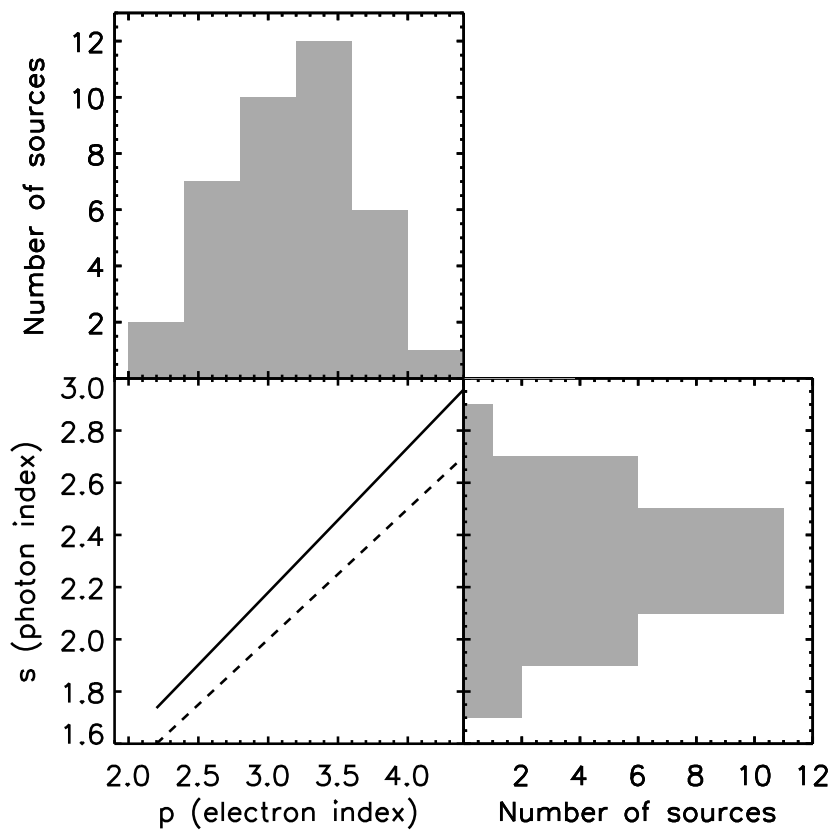}
\caption{Lower left panel: 
the electron spectral index $p$ inferred from the inverse Compton scattering
model as a function of the photon index $s$ 
for the KN (solid line) and  the Thomson (broken line) 
cases. Lower right panel: the distribution of the observed  photon index $s$
for blazars with strong emission lines taken from the third \egret\ catalog
\protect\citep{hartman99}. Upper panel: 
the distribution of the 
electron indices $p$ implied by the \egret\ measurements, taking account of
the individual source redshifts.}
\label{fig2}
\end{figure}

If the \egret-observed flux were due to Thomson scattering, 
the observed photon index $s=\alpha+1$ of a blazar would relate
to the index $p$ of the electron power law via $p=2s-1$. However, 
since KN effects steepen the \egret\ spectrum, application of this relation
 results in an electron index that is too steep, as can be seen 
 in the lower left
panel of figure \ref{fig2}, where  we plot the KN corrected photon index 
(solid line) as well as the relation $s=(p+1)/2$ (dashed line). In this figure,
the seed photon energy is $\epsilon_0=5\times 10^{-5}$,
and the two-point spectral index is calculated 
here and throughout the paper
using the luminosity 
of the model spectrum
at $100 $ MeV and $10  $ GeV.
The maximum electron Lorentz factor is 
assumed to be $\gamma_2{\cal D}>1/\epsilon_0$, since the \egret\  
observations show no evidence of a cut-off.
Under this 
assumption the observed photon index is not a function of $\gamma_2{\cal D}$.
For  photon indices  $s$ around $2.2$ the electron index that would result
from Thomson scattering is steeper than the KN case 
by about $0.3$. 
In the lower right panel of figure \ref{fig2} we show a histogram of the 
observed photon indices of all  blazars  in the third
\egret\ catalog \citep{hartman99}, except for sources with  
errors greater than $0.3$ in
their photon index determination  and also excluding
the three X-ray selected BL Lacertae
(MKN 421, MKN 501, PKS 2155-304),
which, because they lack 
line emission in their spectra,
are believed to be predominantly SSC emitters, 
 
We see that the distribution peaks  at  $s\approx 2.3$ (a careful calculation 
for all the broad line emitting blazars gives $s=2.20 \pm 0.05$ \citep{mukherjee97}). For each  of these  sources we calculate
the electron index $p$ that would result in the observed photon index $s$, 
taking into account the correction due to the redshift of the source, and
plot the results in the upper left panel of figure \ref{fig2}.
The results cluster in a wide interval around $p=3.2$ that includes the 
value $p=3.0$.
These values are expected from electron
distributions with indices $p=2.2$ and $p=2.0$, respectively, that steepen by
$\Delta p=1$ above $\gamma_b$. Since the \egret\ range is at energies
higher than the peak of the spectral energy distribution, the observed
photon indices are compatible with 
acceleration at both relativistic \citep{kirk00, achterbergetal01} 
and non-relativistic shocks, which predict $p=2.23$ and 
$p=2$ respectively for the uncooled electron distributions. 
Unfortunately, however, the scatter in the observed indices is 
of the same order as the difference between the predictions, so that the
data do not allow us to  distinguish
between these two cases.

\begin{figure}
\epsscale{0.75}
\plotone{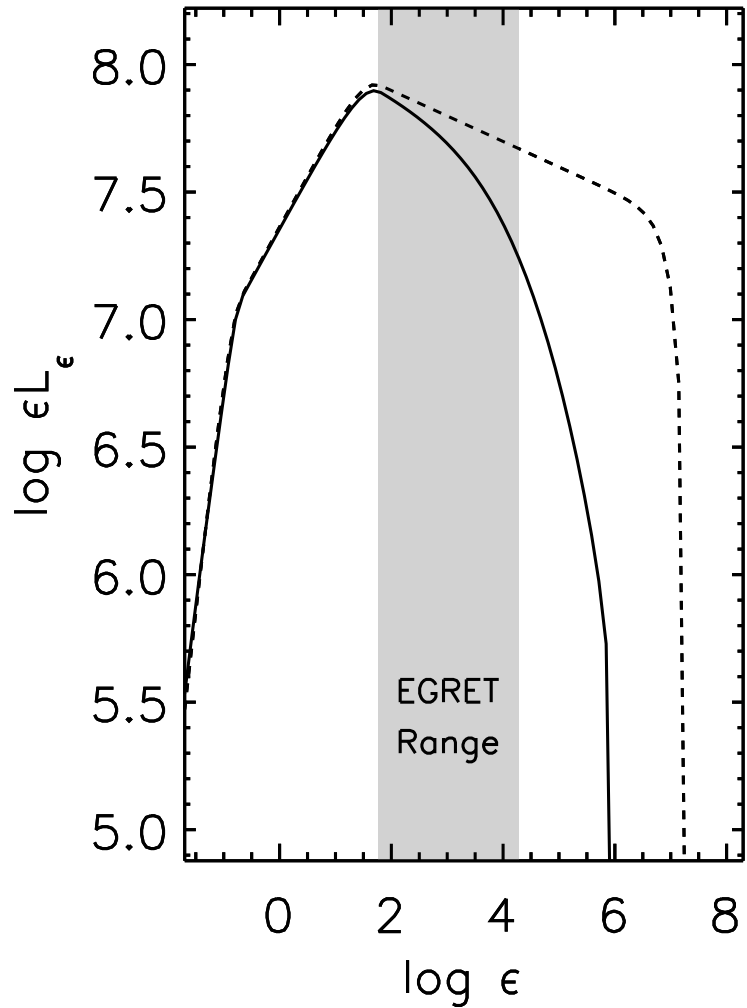}
\caption{ The observed  energy distribution due to inverse Compton 
scattering  for both 
the KN (solid line)
and Thomson treatment (broken line) for  a blob of plasma  that moves with a 
Lorentz factor $\Gamma=10$ through an isotropic monoenergetic photon field.
The electrons in the blob frame are characterized by an isotropic broken power law distribution
$n(\gamma)\propto \gamma^{-p}$, with 
$p=2.2$ for $10 \leq \gamma \leq 2\times 10^2$, and 
$p=3.2$ for $2 \times 10^2 \leq \gamma \leq 10^5$.
We plot, in normalized units,  
the energy distribution due to inverse Compton 
scattering observed at an angle $\theta=1/\Gamma$.
The shaded area corresponds to the \egret\ range of 
observation.}
\label{fig3}
\end{figure}

Simple electron cooling considerations predict a spectral break of 
$\Delta\alpha=0.5$
in the transition before and after the peak energy $\epsilon_{peak}$ of the
observed energy distribution, as a result of  the change in the electron
index  $\Delta p =1$ for synchrotron dominated cooling.
This appears to conflict with the combined \osse, \comptel\ and
\egret\ measurements of some blazars, which found 
 spectral breaks $\Delta\alpha > 0.5 $ 
\citep{mcnaron95,collmar97}.
However, spectral breaks of $\Delta\alpha>0.5$ 
between the \comptel\ and \egret\ ranges
are produced naturally for sources peaking
at MeV energies, since the \egret\  spectrum is softened
by KN effects. We demonstrate this in figure \ref{fig3}, where we plot 
the spectral energy distribution due to inverse Compton scattering of 
optical seed photons by a broken power law electron distribution for both
the KN (solid line) and Thomson calculation (broken line).
The spectrum  below the peak has  a photon  index 
$s=(p+1)/2=1.6$, since below the peak $p=2.2$. Above the peak $p=3.2$ and
the Thomson spectrum has a a photon  index  $s=2.1$, resulting in a 
break $\Delta\alpha=0.5$. The KN spectrum above the peak is steeper,
and the two-point spectral index is calculated to be $s=2.30$, which results
in a spectral break $\Delta\alpha=0.70$. 

\section{Conclusions and discussion\label{discussion}}

In this paper we 
generalize the calculation (D95) of the beaming pattern of inverse Compton
radiation to the KN regime. In terms of the electron index, $p$, 
the beaming factor is 
${\cal D}^{3+p}$, for arbitrary photon energy. The KN-limited 
spectrum produced by a power law
electron distribution is  curved, and the beaming pattern is not a simple
function of 
the local spectral index as in the case of synchrotron emission or SSC.
KN effects further limit the location of the peak of the spectral energy
distribution to
$\epsilon_{peak}\lesssim 1/\epsilon_0$, where $ \epsilon_0$ is the
seed photon energy.
This peak energy is independent of the Doppler factor ${\cal D}$ and the 
maximum Lorentz factor $\gamma_2$ of the electron distribution provided
the scattering is KN limited ($\gamma_2{\cal D}\epsilon_0 \gg 1$), but
does depend on the electron index, with $\epsilon_{peak}$ shifting to lower
energies as the electron distribution becomes steeper.
If the GeV spectrum of the \egret\  blazars with strong emission lines  
is due to external Compton scattering
of optical seed photons
($\epsilon_0 \approx 5\times 10^{-6}$),  this component of the emission
will steepen further due to  KN effects  at energies above  $ 1/\epsilon_0$,
corresponding to   energies $\gtrsim 0.5 $ TeV. Thus,  even in the absence of 
absorption due to the extragalactic IR background, the TeV flux of this EC component is negligibly small.

In many cases, the spectra of blazars detected by \egret\ and observed by
\osse\ and \comptel\ appear to peak in the region $1$ -- $100$ MeV. 
Assuming that this
emission is due to inverse Compton scattering of optical photons 
from the broad line region, the Thomson approximation 
is quite adequate at the peak energy. In this case, there should be a change
in the spectral index measured on each side of the peak of $\Delta\alpha=0.5$,
brought about by cooling via synchrotron or inverse Compton emission.
In several cases, however, it appears that the spectra index determined from
\osse\ and \comptel\ measurements is harder than that in the \egret-band by
$\Delta\alpha>0.5$. This apparent difficulty has 
led to the development of models in which the 
spectral break arises not from cooling, but 
from the differential absorption by pair production at
the edge of a gamma-ray photosphere 
\citep{blandfordlevinson95,marcowithhenripelletier95}.

Our results indicate, however, that the Thomson approximation is not adequate
to describe the spectrum in the \egret\ range. 
Assuming that synchrotron
radiation is the dominant cooling mechanism, we have shown that the spectral
break between the \osse/\comptel\ and \egret\ ranges can substantially exceed 0.5,
without invoking absorption due to pair production.
Our computation assumes that the dominant cooling mechanism is synchrotron
radiation. This is reasonable because the inverse Compton 
flux  $f_{IC}$ is rarely  observed to exceed
the synchrotron flux $f_S$ by more than a factor of 10. 
Application of the beaming formula with $\alpha=1$ implies for the ratio of 
the energy density $U_E$ in target photons to the energy density 
$U_B$ in the magnetic field 
$U_E/U_B=(f_{IC}/f_S)/{\cal D}^2$. For ${\cal D}=10$ we have
$ U_E/U_B \lesssim 0.1$, so that synchrotron cooling dominates.
We note that the 
issue of the  large   ($\Delta \alpha >0.5$) spectral breaks in MeV blazars 
will be accessible to detailed observation following the launch of 
{\it INTEGRAL}.

The spectral softening in the \egret-band brought about by KN effects 
implies underlying electron spectra which are $\sim0.3$ harder than would be
expected from a naive application of the formula $p=(s-1)/2$. As a result, 
the electron indices cluster
around $\approx 3.2$, which, allowing for cooling, corresponds to 
$p\approx 2.2$ below the break
 at $\gamma_b$.  This value is compatible with the predictions
of shock acceleration theory at both at relativistic ($p=2.23$) 
and nonrelativistic
shocks ($p=2.0$). However, the quality of the data do not permit us to
distinguish between these two possibilities.

The association of the peak of the spectral energy distribution with
$\gamma_b$ does not apply to all blazars.
In X-ray bright objects, which include the TeV-emitting sources,
the synchrotron  spectral energy distribution
 is characterized by a sharp decline above the peak, as would be produced
 by a cut-off in the electron energy distribution, rather than a cooling
 break. Observations in the X-ray band of
\lq hard lags\rq\ (hard X-rays
 trailing in time
behind soft X-rays) in  MKN~421
\citep{maraschi99} suggest that at the peak energy
the acceleration and loss time scales of the radiating particles
are comparable 
\citep{kirk97}. The corresponding 
electron spectrum is hard: $p<2$.
This difference in electron spectrum may reflect
a fundamental difference in the underlying particle acceleration 
mechanism in these two distinct classes of objects.

\acknowledgments This work  was supported by the European Union
TMR programme under contract FMRX-CT98-0168.

\newcommand{\eqb}{\begin{eqnarray}}
\newcommand{\eqe}{\end{eqnarray}}
\appendix

\section{The effective observed volume}

Consider a blob containing a uniform distribution of electrons. In the rest
frame of the blob, the differential number of electrons in interval $d\gamma'$
moving within the solid angle $\Omega'$ is
\eqb
{dN'(\gamma')\over d\gamma'd\Omega'}&=&\int_{\rm blob}d^3x' n'(\gamma')
\nonumber\\
&=&V n'(\gamma')
\eqe
where $V$ is the volume of the blob. 
In the lab.\ frame the blob moves along the $x$-axis at speed $c\beta$, and
the direction ${\bf\hat{n}}$ to the observer makes an angle $\theta$ with this
axis. The radiation flux $F_\nu$ 
observed is proportional to the product of the 
single particle emissivity
and the electron distribution integrated over the emitting volume
{\em at the retarded
  time}. This can be written
\eqb
F_\nu&\propto&\int_{\rm blob}d^3x\int_{-\infty}^{+\infty}dt n(\gamma,\mu)
\delta\left(t-{\bf x}\cdot{\bf\hat{n}}/c\right)
\eqe
Apart from the argument of the 
delta function, the only dependence of the integrand on $t$ arises from the 
limits of the integration over the moving blob, so that
\eqb
F_\nu&\propto&n(\gamma,\mu)V_{\rm eff} =N_{\rm eff}(\gamma,\mu)
\eqe
where
\eqb
V_{\rm eff}&=&
\int_{\rm blob}d^3x\int_{-\infty}^{+\infty} dt 
\delta\left(t-{\bf x}\cdot{\bf\hat{n}}/c\right)
\label{vobsdef}
\eqe
is the effective observed volume used in equation (\ref{eq:ele}).
To relate this to $V$, we 
rewrite Eq.~(\ref{vobsdef}) in terms of the spatial coordinates in the blob
frame, keeping the time $t$ in the lab.\ frame. Since $dx=dx'/\Gamma$,
$dy'=dy$ and $dz'=dz$ we have:
\eqb
V_{\rm eff}&=&
{1\over\Gamma}\int_{\rm blob}d^3x'\int_{-\infty}^{+\infty} dt 
\delta\left(t-{\bf x}\cdot{\bf\hat{n}}/c\right)
\nonumber\\
\noalign{\hbox{which becomes, using $x'=\Gamma(x-c\beta t),$}} 
V_{\rm eff}&=&
{1\over\Gamma}\int_{\rm blob}d^3x'\int_{-\infty}^{+\infty} dt 
\delta\left(t-\mu \beta t -\mu x'/(c\Gamma) -\sqrt{1-\mu^2}y'/c\right)
\eqe
where we assume, without loss of generality, that ${\bf\hat{n}}$ lies in the 
$x$-$y$ plane. Performing the integral over $t$ now yields
\eqb
V_{\rm eff}&=&{1\over\Gamma(1-\mu\beta)}\int_{\rm blob}d^3x'
\nonumber\\
&=&V{\cal D}
\eqe

\clearpage

\clearpage

\clearpage

\end{document}